\documentclass[10pt, conference, compsocconf]{IEEEtran}
\ifCLASSINFOpdf
\else
\fi
\usepackage{algorithmic}
\usepackage{algorithm}
\usepackage{url}
\usepackage{cite}
\usepackage[pdftex]{graphicx}
\begin{document}
%\title{Sunflower: Lightweight Coordination for Accelerating Request Response in Distributed Latency-critical Services}
%\title{Sunflower: Lightweight Scheduling for Accelerating Request Response in Latency-critical Datacenter Services}
%\title{Sunflower: Lightweight Scheduling for Reducing Tail Latency in Large-scale Data Center Services}
%\title{Sunflower: Predictive Resource Reassignment for Reducing Tail Latency in Large-scale Data Center Services}
\title{PCS: Predictive Component-level Scheduling for Reducing Tail Latency in Cloud Online Services}


\author{
\IEEEauthorblockN{Rui Han, Junwei Wang, Siguang Huang, Chenrong Shao, Shulin Zhan, Jianfeng Zhan}
\IEEEauthorblockA{Institute Of Computing Technology,\\
Chinese Academy of Sciences\\
Beijing, China\\
{hanrui,wangjunwei,huangsiguang,shaochenrong,zhanshulin,zhanjianfeng}@ict.ac.cn}
\and
\IEEEauthorblockN{Jose Luis Vazquez-Poletti}
\IEEEauthorblockA{Facultad de Informatica,\\
Universidad Complutense de Madrid\\
Madrid, Spain\\
jlvazquez@fdi.ucm.es}
}

\maketitle

%Each component has its own mix of co-located jobs and the workload heterogeneity of batch jobs results in considerably different performance interferences, and thus the component's latency variability.

\begin{abstract}
Modern latency-critical online services often rely on composing results from a large number of server components. Hence the tail latency (e.g. the 99th percentile of response time), rather than the average, of these components determines the overall service performance. When hosted on a cloud environment, the components of a service typically co-locate with short batch jobs to increase machine utilizations, and share and contend resources such as caches and I/O bandwidths with them. The highly dynamic nature of batch jobs in terms of their workload types and input sizes causes continuously changing performance interference to individual components, hence leading to their latency variability and high tail latency.
However, existing techniques either ignore such fine-grained component latency variability when managing service performance, or rely on executing redundant requests to reduce the tail latency, which adversely deteriorate the service performance when load gets heavier.
In this paper, we propose PCS, a predictive and component-level scheduling framework to reduce tail latency for large-scale, parallel online services. It uses an analytical performance model to simultaneously predict the component latency and the overall service performance on different nodes. Based on the predicted performance, the scheduler identifies straggling components and conducts near-optimal component-node allocations to adapt to the changing performance interferences from batch jobs. We demonstrate that, using realistic workloads, the proposed scheduler reduces the component tail latency by an average of 67.05\% and the average overall service latency by 64.16\% compared with the state-of-the-art techniques on reducing tail latency.
\end{abstract}


% \fi
\IEEEpeerreviewmaketitle

\section{Introduction}

\begin{figure}
\centering
  \includegraphics[scale=0.44]{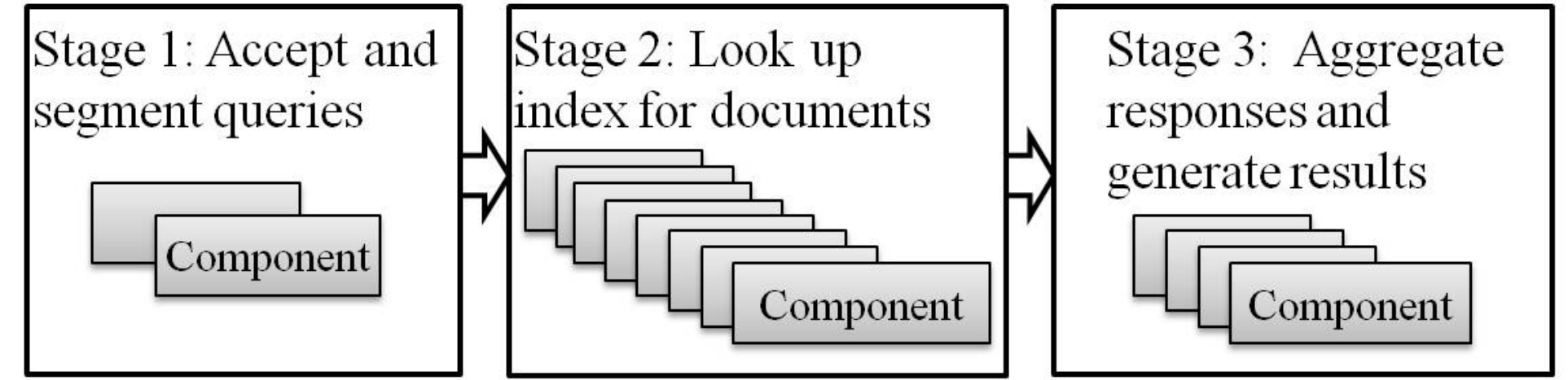}\\
  \caption{An example of Nutch search engine}
  \label{Fig: SearchEngine}
\end{figure}

Providing fluid responsiveness to user requests is essential for online services: their potential profits are proportional to service latency (i.e. request response time including both the request queueing delay and the time of being processed) \cite{jalaparti2013speeding,tailatScale}. In large online services such as search engines, e-commerce sites, and social networks, the processing of incoming requests consists of several sequential stages, where each stage composes responses parallelized across hundreds or thousands of server \emph{components}. Hence the tail (e.g. the 99th percentile) of these components' latencies, rather than the average, determines the overall service performance \cite{tailatScale,kapoor2012chronos}.
For example, Figure \ref{Fig: SearchEngine} shows an example Nutch search engine \cite{nutchsearch} with three stages. Suppose that at stage 2, the request processing is parallelized into 100 components, in which 99 components can respond in 10ms but only one component gets a slow response of 1 second, the overall service performance is deteriorated by this straggling component and hence providing slow responsiveness of 1 second.

%each component can typically respond in 10ms but has a 99th-percentile latency of one second. This small probability of slow request processing is significantly amplified by the service scale and 63.40\% of the requests will encounter at least one slow component and hence take slow responses of more than one second.

%are required to serve numerous end users simultaneously while provide fluid responsiveness to most of the users .

In modern cloud data centers and warehouse-scale computers, it is critical to improve machine utilizations by co-locating long-running online services and offline batch jobs (e.g. Hadoop \cite{HadoopWebsite} and Spark \cite{SparkWebsite} analytics jobs) on the same node (physical machine), while still keeping the overall latency of online services at a satisfactory level \cite{reiss2012heterogeneity,yang2013bubble}.
Although the components of a service are typically hosted on dedicated environments such as Xen virtual machines (VMs) or LinuX Containers (LXCs), these components still share and contend resources such as processing units, caches and I/O bandwidths with their co-running batch jobs on the same node, hence inevitably suffer from performance interference. Workload traces from Google \cite{reiss2012heterogeneity} and Facebook \cite{chen2012interactive} show that small batch jobs form a majority (over 90\%) of all jobs in their data center workloads. For example, approximately 50\% of Google jobs complete in 10 minutes and 94\% of them complete within 3 hours.
These short-term batch jobs have various workload types (e.g. CPU and I/O intensive workloads) and input data sizes (e.g. ranging from KB to GB), thus causing continuously changing performance interferences to their co-located components. This results in the \textbf{component latency variability}, which can be explained from two aspects: (a) each component's latency (performance) varies over time, and (b) components hosted on different nodes have different changes in their latencies, hence causing high tail latency in individual components of the service.

%for each component over time and such variability varies across components hosted on different nodes, hence leads to high tail latency in individual components of the service.

%components hosted on different nodes have distinct latency variability.

%heterogenous and dynamic contentions to the shared resources due to their various workload types (e.g. CPU and I/O intensive workloads), changing job arrival rates and input data sizes, which range from KB to GB. The performance interferences incurred by these short batch jobs, therefore, differ considerably and vary over time across components hosted on different nodes.

%The \textbf{component latency variability} caused by such performance interference gives rise to a key practical challenge to be addressed to provide short latency tail for all components of the service.

%in the existence of heterogenous batch workloads
%The performance of the component is also impacted by background hardware/software activities in its host machine \cite{tailatScale}.

%one major reason that hinders the consolidation of these applications

Many existing techniques have been developed to guarantee the performance of latency-critical services by mitigating the performance interference due to resource sharing and contention \cite{kasture2014ubik,xu2013bobtail,ahn2012dynamic,xu2010mitigating}. However, these techniques only manage service performance at the coarse granularity of the entire application, ignoring fine-grained component latency variability that may come to dominate service performance at large scale. Moreover, state-of-the-art techniques reduce tail latency via request redundancy. They either create replicas for all the requests \cite{vulimiri2012more,ananthanarayanan2013effective,stewart2013zoolander} or reissue slow requests' replicas to a different component \cite{jalaparti2013speeding, tailatScale}, and then use the quickest replica. Although these techniques work well under light load, they adversely deteriorate the service performance when load gets heavier \cite{shah2013redundant}.

%--> as the size and complexity of the system scales up or as overall use increases.

%many generic infrastructure resources such as on-chip caches, data prefetchers, the memory bus and controllers, and disk and network bandwidths are still shared and contended by applications co-scheduled in the same machine

%Queueing or scale, few slow storage accesses per request can raise response times a lot, making the whole service less usable and hurting profits

In this paper, we propose a new component-level service scheduler that dynamically schedules the components of a service to appropriate nodes with the assistance of cost-effective online monitors. Compared to existing latency reduction techniques, the proposed scheduler applies an analytic performance model to predict the latencies of all components and their impact on the overall service performance, and then formulates the scheduling decisions based on the predicted performance. The performance model also dynamically updates the prediction results at each scheduling interval by collecting the latest resource contention information during the service execution, thus allowing the scheduler to adapt to changes in performance interference. The concrete contributions of this work are as follows:
\begin{itemize}
\item  We build a flexible analytic performance model to accurately predict the performance of an online service. The basic model comprehensively covers some of the most representative shared resources that are likely to incur contentions and predicts each component's service time on different nodes by taking the resource contention and performance interference into consideration. The extended model further considers the request queueing delay and estimates the latency of the whole service based on its implementation topology. We show that the proposed model can predict the latency with an average error of 2.68\%.
\item  Based on the performance model, we present a framework for component-level scheduling. At each scheduling interval, our approach efficiently identifies the straggling components of a service such that the migration of these components brings the maximum reduction in the overall service latency. The effectiveness of the proposed approach is evaluated using comparative experiments on a variety of realistic workloads publicly available from the BigDataBench suite \cite{opensourceBigDataBench}. The experiment results in a 100-machine cluster demonstrate that compared with the state-of-the-art techniques on mitigating tail latency, our approach reduces components' 99th percentile latency by an average of 67.05\% and the average overall service latency by 64.16\%.
\end{itemize}
The remainder of this paper is organized as follows: Section \ref{Section: Background} introduces the background information. Section \ref{Section: Overview of the framework} gives an overview of the proposed scheduling framework. Section \ref{Section: Performance predictor} presents the performance model and Section \ref{Section: Algorithm} explains the scheduling algorithm. Section \ref{Section: Evaluation} evaluates the proposed approach. Section \ref{Section:Related Work} presents the related work, and finally, Sections \ref{Section: Conclusion} summarizes the work.

%Secondly and more importantly, from the \textbf{whole service}'s perspective, the variability in servers' response time, in particular straggler servers with longer response time, is significantly amplified for large services distributed in hundreds of servers, which incurs violations of tail latency.

\section{Background} \label{Section: Background}
\subsection{Sources of component latency variability} \label{Section:Sources of tail latency}

\emph{Resource sharing and contention}. When deploying an online service on a cloud platform, the performance interference due to the co-located batch jobs' resource contention is often regarded as a major cause of a component's service time variability \cite{tailatScale,leverich2014reconciling}. Some system activities including hardware activities (such as garbage collections of storage devices and energy management behaviors) and software activities (such as kernel daemons and system maintenance) also influence the component's service time.

\emph{Queueing delay}. The component's service time variability is significantly amplified in the request queueing delay when considering different request arrival rates. Hence the variability of service time and queueing delay work together to cause large latency variability in individual components.

\subsection{Dynamic performance interference of batch jobs} \label{Section:Managing data center applications}
%\subsection{Managing workload heterogeneity and dynamicity in data center applications}
%\subsection{Workload heterogeneity and dynamicity in batch applications}

%Today's data centers are designed to execute a mix of latency-critical services and batch jobs, providing low latencies for services while using underutilized resources to execute batch jobs so as to maximize resource utilizations.

The dynamic performance interference of batch jobs are caused by their short running periods and continually changing workload characteristics, which can be explained in two aspects.

%A key challenge of addressing performance interferences from batch jobs is to understand their workload characteristics.
%managing such mixed workloads is to understand their workload characteristics of a wide variety of batch jobs and their
%impact on resource usage and contention. This workload heterogeneity can be typically explained in two aspects.

\emph{Workload type}. It has twofold meanings: (i) \emph{Computation semantics}. Batch jobs with different computation semantics (i.e. source codes) may have different resource demands. For example, Sort is an I/O-intensive workload, Bayes classification is a CPU-intensive workload with dominated floating point operations, and Page Index has similar demands for CPU and I/O resources. (ii) \emph{Software stacks}. Model software stacks such as Hadoop and Spark usually provide rich libraries to facilitate development of new applications, and allow a programmer focus on writing a few lines of codes to implement an application. Hence a batch job of the same computation semantic may have considerably different resource demands when implemented with different software stacks \cite{jia2014characterizing}. For example, Hadoop Bayes is a CPU-intensive workload but Spark Bayes is an I/O-intensive workload.

\emph{Input data size}. The resource demand of a job varies when it processes different input data sizes. For example, when running on a 12-core Xeon E5635 processor, the CPU utilizations of the WordCount workload are 31\%, 61\%, and 79\% when its input data sizes are 500MB, 2GB, and 8GB, respectively.

\section{Overview of the framework} \label{Section: Overview of the framework}

\begin{figure}
\centering
  \includegraphics[scale=0.45]{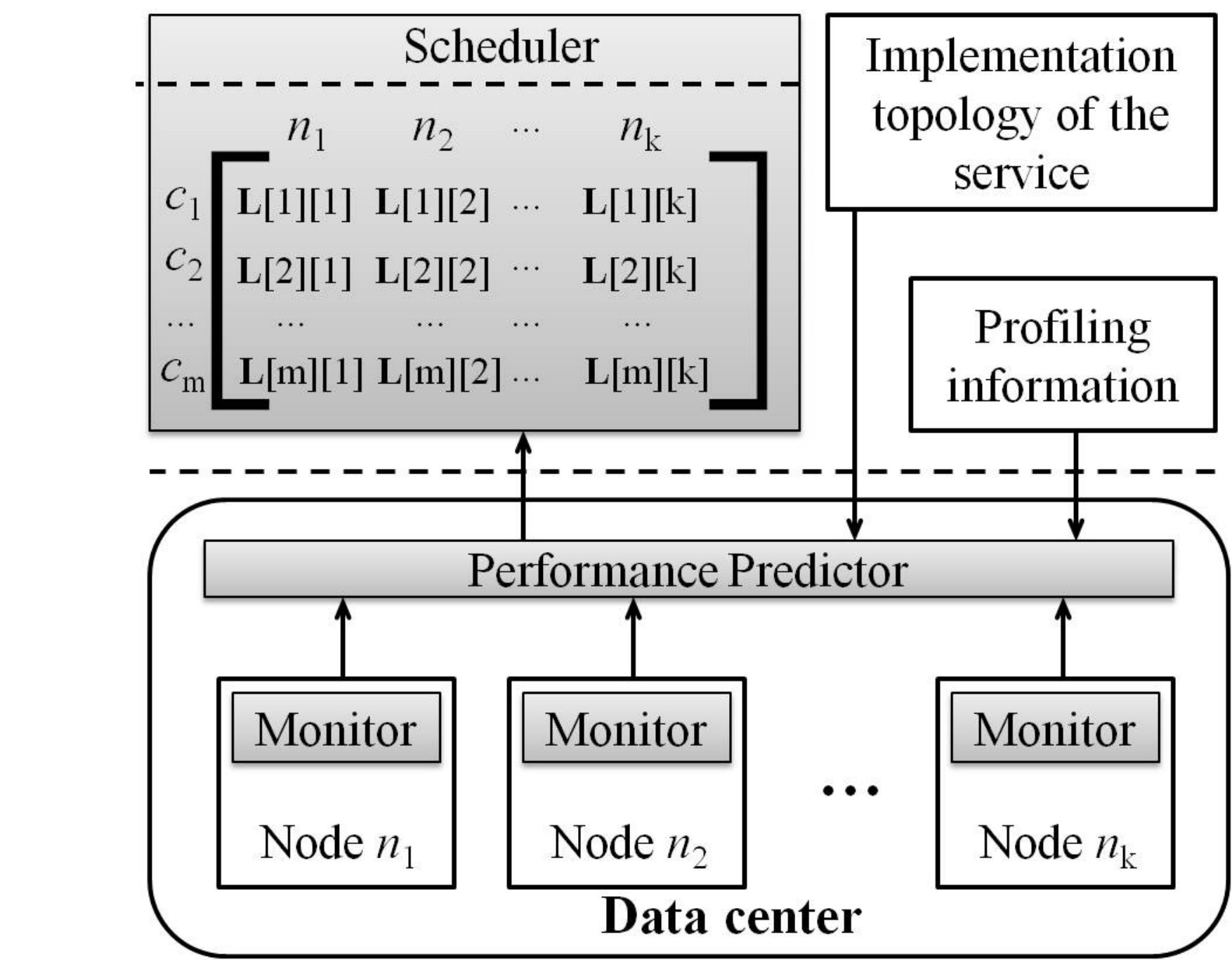}\\
  \caption{The overview of the framework}
  \label{Fig: Overview}
\end{figure}

As shown in Figure \ref{Fig: Overview}, the proposed framework for predictive component-level scheduling consists of three modules: the on-line monitors, the performance predictor and the scheduling heuristic.

The \emph{on-line monitor} continuously detects two types of information in a running service, whose components are distributed on $k$ nodes of a data center. The first type of information represents the service's workload status, i.e. its request arrival rate. The second type of information reflects the resource contention information of each component due to its co-located programs on the same node. Specifically, the monitor obtains the request arrival rate by profiling service's running logs, collects system-level contention information (e.g. core usage and I/O bandwidths) by accessing the proc file system, and profiles micro-architectural contention information (e.g. shared cache misses) using hardware performance counters for Linux 2.6+ based systems. In our monitor, Perf \cite{PerfWebsite} is used to profile physical machines and Oprofile \cite{OprofileWebsite} is used to profile VMs.

At the end of each scheduling interval, the \emph{performance predictor} collects the monitored information and predicts the component's latency on all $k$ nodes. This predictor also estimates the impact of individual component latencies on the performance of the whole service based on its implementation topology, and organizes the predicted values as a performance matrix. Using this matrix, the \emph{data center scheduler} applies the scheduling algorithm to identify the straggling components and enforces the appropriate node assignment of the components for the next interval. Consequently, the framework is able to dynamically and efficiently adapt to component latency variability.

Note that the proposed scheduling algorithm is not intended to replace, but rather complement the existing scaling or resource provisioning techniques (e.g. reactive scaling up \cite{han2012lightweight} or prediction-based resource provisioning \cite{calheiros2011virtual,han2014enabling} approaches) for multi-stage online services. Specifically, the component-level scheduling is enforced only after the machines have been allocated to the service. At each scheduling interval, the component-node allocation can be conducted by calling the deployment APIs offered by existing distributed realtime computation systems such as Storm \cite{StormWebsite} and Drill \cite{DrillWebsite} to migrate the components to the available machines (e.g. VMs or LXCs) on the scheduled nodes. Note also that although this component-node allocation can be enforced by directly migrating the machines to the nodes, we prefer the former solution as it produces lower overheads on scheduling.

%in order to redeploy the components on the available machines of the service

\section{Performance predictor} \label{Section: Performance predictor}

%Given workload complexity and unpredictability of these causes, they argue that completely eliminating tail latencies of datacenter applications is impractical.
%So not possible to search all possible solutions

Predicting a component's latency when running on different nodes is the key step to detect straggling components in a service. This requires the performance predictor to consider all causes of latency variability discussed in Section \ref{Section:Sources of tail latency}. In the presence of fine-grained heterogeneity of resource contentions on each component, the basic performance predictor is responsible for collecting the information of resource sharing and contention, and predicting the impact on individual component's performance (Section \ref{Section: Basic performance model}). The extended performance model further estimates the component's latency by taking the current request arrival rate into account, and calculates the overall service latency based on the service implementation topology (Section \ref{Section: Extended Performance Model}). With these two models, the performance predictor finally exposes the component latency variability to the scheduler as a performance matrix of reduced overall service latencies (Section \ref{Section: Matrix Construction}). Table \ref{table: Table of notations} lists all notations.

\begin{table}[h!]
  \caption{Table of notations}
  \centering
  \begin{tabular}{|l|l|}
    \hline
    \textbf{Symbol} & \textbf{Meaning} \\
    \hline
    $n$ & A node \\
    \hline
    $c$ & A component belonging to a service\\
    \hline
    $x$ & $c$'s service time \\
    \hline
    $sr$ & One type of shared resources \\
    \hline
    $U_{sr}$ & $sr$'s resource contention information \\
    \hline
    $\textbf{U}$ & The contention vector consists of contention \\
    &  information of all shared resources\\
    \hline
    $RG(U_{sr})$ & A basic regression model \\
    \hline
    $RG_{ST}(\textbf{U})$ & A combined regression model representing\\
    &the predicted service time\\
    \hline
    $l$ & $c$'s latency \\
%    \hline
%    $\triangle l$ & $c$'s reduced latency in migration \\
    \hline
    $l_{stage}$ & A stage's latency \\
    \hline
    $l_{overall}$ & The overall service latency \\
    \hline
    $\textbf{L}$ & The matrix of the reduced overall service latency \\
    \hline
    $\textbf{L}[i][j]$ & An entry of \textbf{L}, which represents the reduced \\
    & overall latency when component $c_{i}$'s is migrated\\
    & from its current node to node $n_{j}$ \\
    \hline
  \end{tabular}
  \label{table: Table of notations}
\end{table}

%Hence when co-locating each of the three applications individually with the Nutch search engine, each batch application has distinct resource interference to its co-runner, as shown in Figure 1(c). <three batch applications, different aspects of resource interferences>.

\subsection{Basic performance model} \label{Section: Basic performance model}

Given a component $c$ hosted on a node $n$, the basic performance model is developed to capture the impact of resource sharing and contention on $c$'s performance and estimate its service time $x$.

Table \ref{table: Usage information of shared resources} lists the contention information of shared resources. The model comprehensively considers both on-chip resources (e.g. shared processing units and caches) and off-chip resources (disk and network bandwidths) contended by different programs on node $n$. In Table \ref{table: Usage information of shared resources}, core usage represents the ratio of time running instructions on the cores (including private cache hits); MPKI represents the number of instruction Misses Per Kilo Instructions of shared caches including last level cache (LLC), instruction Translation Lookaside Buffer (TLB/ITLB), and data TLB (DTLB). MPKI thus indicates the stalled cycles due to cache contention. Note that the \emph{contention} of these resources comes from $c$'s co-running programs within the same service or across other applications, and node $n$'s hardware/software activities.

%instructions running on user and kernel modes

%represents the time running instructions on the core (and hitting in private caches)
%Toff_core represents the stalled cycles due to memory accesses (including shared caches)

%(L1, L2 are typically private caches)

%a component's shared resources, which consists of two parts: the resource usage of the component's co-running applications in the same node and other resource usage of this node (e.g. resource usage of background daemons).

%Note that the contention of these resources come from $c$'s co-running servers within the same service or across different applications, and node $n$'s hardware/software activities.

%the unit for $U_{diskBW}$ and $U_{networkBW}$ is MB per second.

\begin{table}[h!]
  \caption{Contention information of shared resources}
  \centering
  \begin{tabular}{|l|l|}
    \hline
    \textbf{Shared resources} & \textbf{Contention information} \\
    \hline
    Floating point and vector  & $U_{core}$=core usage\\
    processing units, pipelines,& \\
    and data prefetchers & \\
    \hline
    LLC, ITLB, DTLB & $U_{cache}$=MPKI\\
    \hline
    Disk bandwidth & $U_{diskBW}$=the amount of \\
    & read/write data per second \\
    \hline
    Network bandwidth & $U_{networkBW}$=the amount \\
    & of send/receive data per second  \\
    \hline
  \end{tabular}
  \label{table: Usage information of shared resources}
\end{table}

Based on the contention information, the basic performance model predicts $c$'s service time $x$ using two steps. The first step employs a regression model to describe the relationship between one contention information and $c$'s service time.
The training of the regression model takes a set of $v$ samples \{({$U_{sr_1}$,$x_{1}$}),...,({$U_{sr_k}$,$x_{v}$})\} as input and outputs a model $RG(U_{sr})$, where $sr \in \{core, cache, diskBW, networkBW\}$ and $U_{sr_i} \in \{U_{core_i}$,$U_{cache_i}$,$U_{diskBW_i}$,$U_{networkBW_i}\}$ ($i$= 1,...,$v$). Hence $c$'s service time $x$ is predicted as $RG(U_{sr})$ when the contention information is $U_{sr}$. The training samples are obtained from profiling runs or historical running logs.

During the training of regression model, the first step also calculates the relevance (i.e. weight $w_{sr}$) between the contention information of shared resource $sr$ and $c$'s service time.
Suppose four regression models ($RG_{U_{core}}$, $RG_{U_{cache}}$, $RG_{U_{diskBW}}$ and $RG_{U_{networkBW}}$) and their weights ($w_{core}$,$w_{cache}$,$w_{diskBW}$ and $w_{networkBW}$) are obtained, the second step predicts $c$'s service time $x$ by producing the final regression model $RG_{ST}(\textbf{U})$ that takes a weighted combination of all the four models:

\begin{equation}
RG_{ST}(\textbf{U})=\frac{\sum_{i=1}^4(w_{sr_i} \times RG_{U_{sr_i}})}{\sum_{i=1}^4 w_{sr_i}}
\label{Equation: regressionModel}
\end{equation}
where the resource contention vector $\textbf{U}=\{U_{core},U_{cache},U_{diskBW},U_{networkBW}\}$.

%In many regression scenarios, this boosting strategy results in significantly improvements in the prediction accuracy \cite{friedman2000additive}.
%Previous studies show that the performance interference of resource sharing and contention can be predicted accurately based on profiling information and history data of previously workloads \cite{delimitrou2013ibench,delimitrou2014quasar, delimitrou2013paragon}.

%Two reasons:
%1) only small profiling --> individual resources; adoboost--> small but high accurate predictions --> only small samples are needed
%2) only idle state, no consideration of queueing delay.
%In the basic performance model, the profiling only needs to be conducted using light workloads such that the service is at idle state,
%Hence the profiling period is short and it can be completed within a few minutes.

%(that is, no consideration of queueing delay in the basic performance model)

%>  is based on the previously proposed interval analysis

% and it represents the resource usage vector of $c$'s shared resources.

%co-location on queuing delay can be thoroughly estimated by running the latency-critical service concurrently with micro-benchmarks that put increasing amounts of pressure on individual resources

%, which consists of two parts: the ideal ideal service time $x_{\phi}$ without any interferences and

\subsection{Extended Performance Model} \label{Section: Extended Performance Model}

%> Explain, adapt to different scenarios, different servers of different values of mean and variance.

%>>> discuss queueing model, no equations

The extended performance model further employs the queueing system to estimate individual component latency under different request arrival rates.
Typically, a queueing system can be described as an $A/X/m$, where $A$ represents the distribution of interarrival time of requests; $X$ denotes the distribution of service time; and $m$ is the number of servers.
The choice of M/G/1 queueing system is based on the assumption that the distribution of interarrival time of incoming requests are determined by a Poisson process (M for Markov); a component is modeled as a server in the queueing system and the distribution of its service time can follow arbitrary distributions (G for General). Let $\lambda$ be the monitored request arrival rate and $\mu$ be the service rate. Let $\bar{x}$ be the mean service time ($\bar{x}$=$1/\mu$), and $var(x)$ be the variance of service time. Component $c$'s expected latency $l$ is calculated as:

\begin{equation}
l=\bar{x} + \frac{\lambda(1+C^{2}_{x})}{2\mu^2(1-\rho)}
\label{Equation: a server's response time}
\end{equation}

where $C^{2}_{x}=\frac{var(x)}{\bar{x}^2}$ is the squared coefficient of variation of service time $x$ and $\rho=\frac{\lambda}{\mu}$ is the server utilization. In many service components, when the service time follows the exponential distribution, that is, the squared coefficient of variation $C^{2}_{x}=1$, the M/G/1 queueing system equals the M/M/1 queueing system and the expected latency $l=\frac{1}{\mu-\lambda}$.
At each scheduling interval, a set of resource contention vectors can be collected for each component. By substituting them into Equation \ref{Equation: regressionModel}, the component's corresponding service time $x$ can be estimated, so its mean and variance can be calculated.

% including both requests' \emph{queueing time} and \emph{the time of being processed}
%In addition, the M/G/1 queueing system supports different service disciplines of requests such as first-in first-out (FIFO), fair queue, and priority queue. These service disciplines correspond to the job scheduling policies in Storm and Spark Streaming frameworks.

Furthermore, the model computes the overall latency of a service based on its implementation topology. In the online services studied in this work, the processing of a request includes several sequential stages, and each stage parallelizes requests across one or multiple components to aggregate their responses. Hence the calculation of an overall service latency consists of two steps. The first step computes the latency of each stage. Suppose a stage consists of $C$ parallel components, its latency is the maximum value of these component latencies:
\begin{equation}
l_{stage}= \max_{1 \leq i \leq C}\{l_{i}\}
\label{Equation: a stage's response time}
\end{equation}
where $l_{i}$ is the latency of component $c_{i}$ ($i$= 1,...,$C$).

Suppose the service consists of $S$ sequential stages, the second steps calculates its overall latency:
\begin{equation}
l_{overall}= \sum_{j=1}^S l_{stage_{j}}
\label{Equation: the overall response time}
\end{equation}
where $l_{stage_j}$ is the latency of the $j$th stage ($j$= 1,...,$S$).
 
\subsection{Performance Matrix} \label{Section: Matrix Construction}

Suppose $m$ components of a service are deployed in $k$ nodes, the $m \times k$ performance matrix $\textbf{L}$ is constructed using components as rows and nodes as columns. An entry $\textbf{L}[i][j]$ denotes the changes in the \emph{overall service latency} $l_{overall}$ when a component $c_{i}$ is migrated from its current node $n_{current}$ to node $n_{j}$ ($1 \leq i \leq m$ and $1 \leq j \leq k$). This migration may influence all $m$ components' contention vectors. For any component $c$ of the service, let its original resource contention vector be $\textbf{U}$ and the updated contention vector after the migration be $\textbf{U}^{'}$. Let the resource contention from $c_{i}$ itself be $\textbf{U}_{c_{i}}$ and the resource consumption from all programs on node $n_j$ be $\textbf{U}_{n_{j}}$.  Four situations needs to be considered when calculating the updated contention vector $\textbf{U}^{'}$, as listed in Table \ref{table: Updated usage vectors in migration}.

\begin{table}[h!]
  \caption{Calculation of the updated contention vector $\textbf{U}^{'}$}
  \centering
  \begin{tabular}{|l|l|}
    \hline
    \textbf{Type of component $c$} & \textbf{Updated contention vector $\textbf{U}^{'}$} \\
    \hline
    $c_{i}$& $\textbf{U}_{n_{j}}$\\
    \hline
    Any component on $n_{current}$ & $\textbf{U}$-$\textbf{U}_{c_{i}}$\\
    \hline
    Any component on $n_{j}$ & $\textbf{U}$+$\textbf{U}_{c_{i}}$\\
    \hline
    Any other component & $\textbf{U}$\\
    \hline
  \end{tabular}
  \label{table: Updated usage vectors in migration}
\end{table}

By substituting $\textbf{U}^{'}$ into Equations \ref{Equation: regressionModel} and \ref{Equation: a server's response time}, $c$'s updated latency $l_i^{'}$ can be calculated. We have: (i) $c_{i}$'s latency $l_i^{'}$ decreases if $n_j$ has lighter resource contention than $n_{current}$; otherwise $l_i^{'}$ increases. (ii) All the components on node $n_{current}$ have decreased latencies because the removal of $c_{i}$ alleviates the resource contention on $n_{current}$. (iii) All the components on node $n_{j}$ have increased latencies because the addition of $c_{i}$ aggravates the resource contention on $n_{j}$. (iv) the latencies of other components keep unchanged. Furthermore, by substituting the updated latencies of all $m$ components into Equations \ref{Equation: a stage's response time} and \ref{Equation: the overall response time}, the updated overall service latency $l_{overall}^{'}$ can be calculated. Let the overall latency be $l_{overall}$ before the migration. The entry $\textbf{L}[i][j]$ can be calculated as:
\begin{equation}
\textbf{L}[i][j] = l_{overall} - l_{overall}^{'}
\label{Equation: changes in the overall response time}
\end{equation}

%change of the overall response time $c_ij$ because of this migration can be calculated.
%Thus, $\triangle x > 0$ means $s$'s service time is increased.

\begin{figure}
\centering
  \includegraphics[scale=0.40]{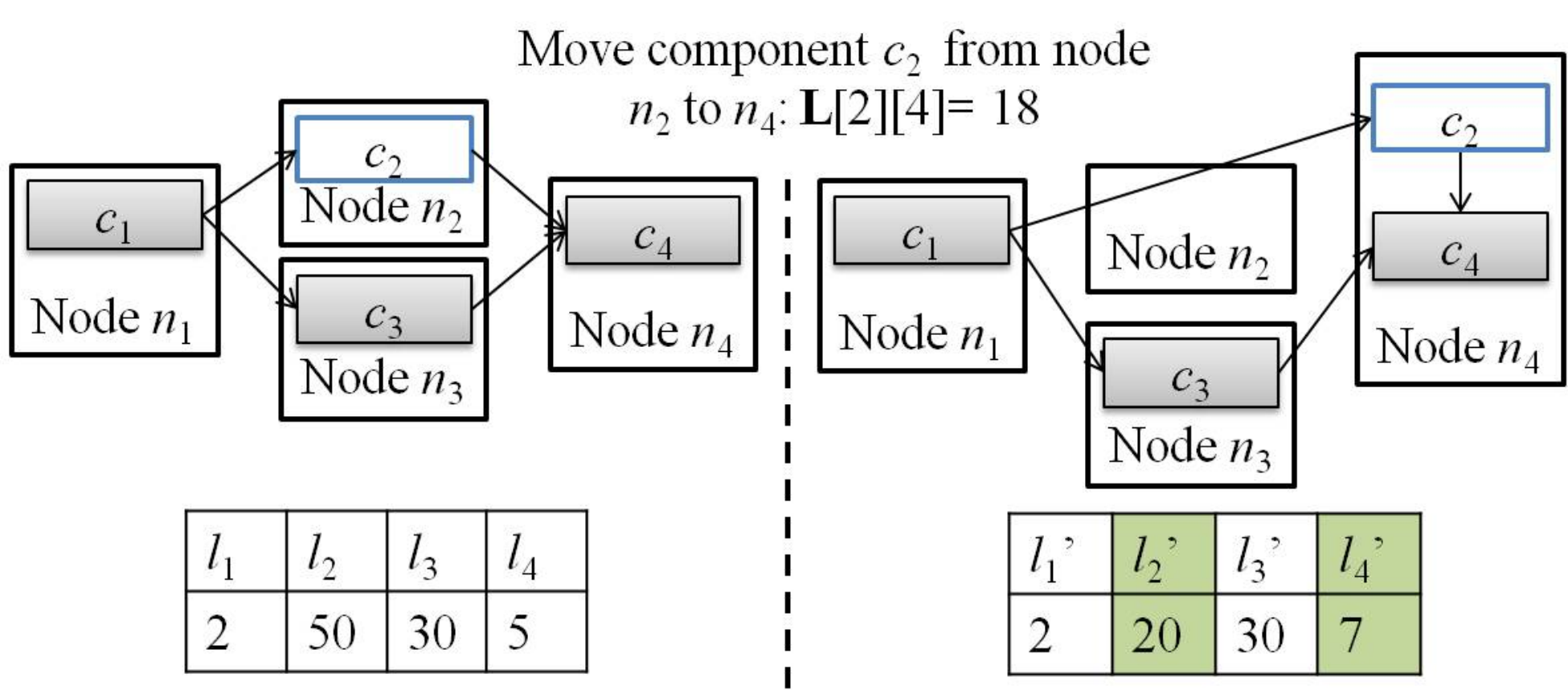}\\
  \caption{An example of entry \textbf{L}[2][4] in the performance matrix}
  \label{Fig: EntryInMatrix}
\end{figure}

Figure \ref{Fig: EntryInMatrix} shows an example service with three stages, where stage 2 is parallelized into two components $c_2$ and $c_3$. After $c_2$ is migrated from node $n_2$ to $n_4$, $c_4$'s latency $l_4$ increases and $c_2$'s latency $l_2$ decreases. By considering all the updated latencies, the overall service latency before and after the migration can be calculated: $l_{overall}$=57ms and $l_{overall}^{'}$=39ms. Hence the reduced latency $\textbf{L}[2][4]$=18ms.

%Replacing into Equations, get the updated total response time, i.e., $c_ij$

\section{The Component-level Scheduling Algorithm} \label{Section: Algorithm}

Based on the performance matrix of a service, the scheduler can conduct component-node allocations to minimize the overall service latency. Let the $m$ components \{$c_1$,...,$c_m$\} be deployed in $k$ nodes \{$n_1$,...,$n_k$\}, a na\"{i}ve approach needs a time complexity $O(k^m)$ to identify the optimal component-node allocation and such exhaustive search is not scalable in practical scenarios. However, performance interference is changing overtime, hence optimizing component-node allocation for a particular dynamic scheduling is not worthwhile.
The proposed approach, therefore, applies a greedy algorithm with polynomial computation complexity and the algorithm has several iterations. At each iteration, the algorithm aims to minimize the overall service latency by evaluating all possible component-node migrations and selecting the migration that would reduce the latency the most.

The pseudocode of the algorithm is presented in Algorithm \ref{SchedulingAlgorithm}. At each scheduling interval, the algorithm first constructs the performance matrix $\textbf{L}$ using the performance mode and the monitoring information (line 2). The initial candidate array $C[N_s]$ takes all $m$ components as its elements (line 3). The scheduling process then iteratively executes under two conditions: (a) $C[N_c]$ is not empty; (b) at least one component in $C[N_c]$ can be migrated (line 5). The second condition indicates that a migration is enforced only when the predicted maximum reduced overall latency $l_{max}$ is larger than a specified threshold $\varepsilon$. This threshold prevents inefficient migrations such that the reduced latency cannot compensate the migration cost. In each loop (line 5 to 15), the algorithm first traverses the matrix $\textbf{L}$ to identify a set $S_L$ of entries with the largest value (line 6). Any of these entries denotes the migration of a component that brings the maximal reduction in the overall service latency. If set $S_L$ contains multiple entries,
the algorithm further searches $S_L$ to find the entry $\textbf{L}[cmax][nDestination]$ representing the migration that brings the largest reduction to the latency of the migrated component itself (line 7). Component $c_{cmax}$ is regarded as the straggling component and it is allocated to node $n_{nDestination}$ (line 11). The migrated component $c_{cmax}$ is then removed from the candidate array $C[N_c]$ and the matrix $\textbf{L}$ is updated after this migration.

%Hence the algorithm traverses the matrix to search the entry $\textbf{L}[cmax][nDestination]$ with the maximal value, and stores the corresponding component-node allocation in array $A[m]$, which represents the deployment plan of the service.

\begin{algorithm}[htbp]
\caption{Predictive Component-level Scheduling}
\label{SchedulingAlgorithm}
\algsetup{
linenosize=\small,
linenodelimiter=.
}
\begin{algorithmic}[1]
%\WHILE{$R_{n_{max}}$>$R_{median}$ and existing running batch applications}
% $C$: the matrix of changed response time;\\\
\REQUIRE $m$: the number of components;\\
$k$: the number of nodes;\\
$N_c$: the number of candidate components to be migrated;\\
$C[N_c]$: the index array of candidate components; \\
$\varepsilon$: the migration threshold; \\
$A[m]$: the component-node allocation array, where $A[i]$ represents the index of the $i$th component's hosting node; \\
$cmax$: the index of the straggling component $c_{cmax}$;\\
$nOrigin$: the index of $c_{cmax}$'s original node;\\
$nDestination$: the index of $c_{cmax}$'s destination node;\\
$l_{i,j}^c$: A component $c$'s reduced latency when migrating from note $n_i$ to $n_j$.\\

\STATE Obtain the monitoring information once every scheduling interval;
\STATE Construct the performance matrix $\textbf{L}$;
\STATE $C[N_c]$=\{$c_1$,...,$c_m$\};
\STATE $l_{\max}$=$\varepsilon$+1;
\WHILE{($C[N_c]$ is not empty and $l_{\max} > \varepsilon$)}
 %   \STATE $l_{\max}$=$\varepsilon$; %// the maximal reduced overall response time
    \STATE Find a set of entries $S_L$ in the performance matrix L with the largest value;
    \STATE Find the entry $\textbf{L}[cmax][nDestination]$ in $S_L$ with the largest value $l_{nOrigin,nDestination}^{c_{cmax}}$;
    %, where $n_i$ and $n_j$ are component $c$'s original and destination nodes, respectively;
 %   \STATE $cmax=i$, $nDestination=j$;
    \STATE $l_{\max}=\textbf{L}[cmax][nDestination]$;
    \IF{($l_{\max} > \varepsilon$)}
        \STATE $nOrigin$ = $A[cmax]$;
        \STATE $A[cmax]=nDestination$; %//Migrate server $s_{smax}$ to $n_{nmax}$
        \STATE Remove $c_{cmax}$ from $C[N_c]$;
        \STATE UpdateMatrix($\textbf{L}$, $C[N_c]$, $A[m]$, $nOrigin$, $nDestination$);
    \ENDIF
\ENDWHILE
\STATE Enforce component-node allocation based on $A[m]$.
\end{algorithmic}
\end{algorithm}

The detailed matrix updating function is given in Algorithm \ref{UpdatingFunction}. The migration of component $c_{cmax}$ from node $n_{nOrigin}$ to $n_{nDestination}$ alleviates the resource contention on $n_{nOrigin}$ but aggravates the resource contention on $n_{nDestination}$, hence has a twofold impact on the predicted reduction of the overall latency for the following migrations. First of all, components to \emph{migrate to} $n_{nOrigin}$ ($n_{nDestination}$) have \emph{increased} (\emph{decreased}) reductions in the overall latency. Hence the entries in the nOriginth and nDestinationth columns should be updated according to Equations \ref{Equation: regressionModel} to \ref{Equation: changes in the overall response time} (line 1 to 5). Secondly, components to \emph{migrate out} of $n_{nOrigin}$ ($n_{nDestination}$) have \emph{decreased} (\emph{increased}) reductions in the overall latency. Each of such components is hosted on either $n_{nOrigin}$ or $n_{nDestination}$ (line 3) and the component corresponds to one row in the matrix, hence the entries in this row should be updated (line 7 to 10). Note that component $c_{cmax}$ is removed from the candidate array $C[N_c]$, so all the entries related to $c_{cmax}$ are not updated.

\begin{algorithm}[htbp]
\caption{UpdateMatrix($\textbf{L}$, $C[N_c]$, $A[m]$, $nOrigin$, $nDestination$)}
\label{UpdatingFunction}
\algsetup{
linenosize=\small,
linenodelimiter=.
}
\begin{algorithmic}[1]
\REQUIRE $N_r$: the number of rows to be updated;\\
$R[N_r]$: the index array of rows; \\
\FOR{($i$=0; $i<m$; $i$++)}
    \STATE Update $\textbf{L}[i][nOrigin]$ and $\textbf{L}[i][nDestination]$;
    \IF{(($A[i]$==$nOrigin$ or $A[i]$==$nDestination$) and $c_i \in C[N_c]$)}
        \STATE Add the $i$th row $r_i$ to $R[N_r]$;
    \ENDIF
\ENDFOR
\FOR {each row $r_j$ in $R[N_s]$}
    \FOR{($v$=0; $v<k$; $v$++)}
        \STATE Update $\textbf{L}[j][v]$;
    \ENDFOR
\ENDFOR
\end{algorithmic}
\end{algorithm}

\begin{figure}
\centering
  \includegraphics[scale=0.45]{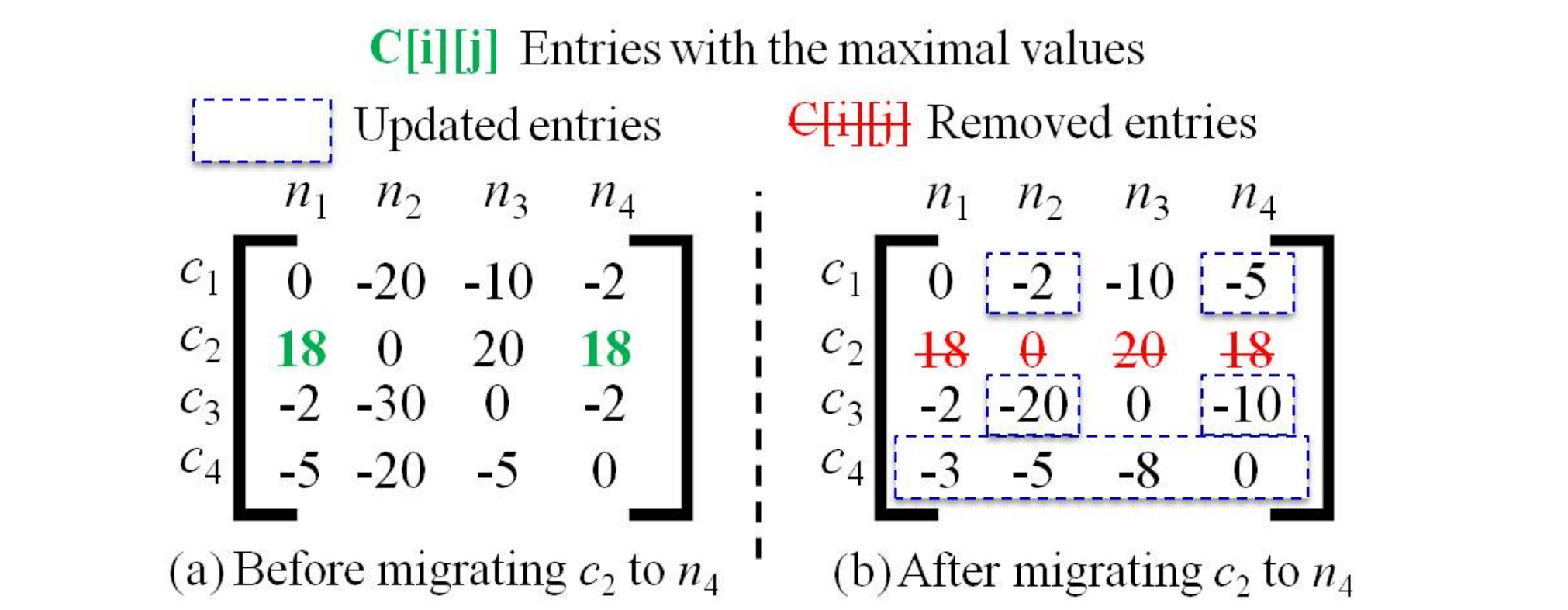}\\
  \caption{An example loop of migrating component $c_2$ to node $n_4$}
  \label{Fig: ServerMigration}
\end{figure}

Figure \ref{Fig: ServerMigration} illustrates an example loop, in which migrating component $c_2$ to either node $n_1$ or $n_4$ can result in the maximal reduction in the overall service latency. At the same time, the reduction in $c_2$'s latency is 20ms when it is migrated to $n_1$ and 30ms when it is migrated to $n_4$, which indicates $c_2$ suffers from less performance interference when it is hosted on $n_4$. Hence the scheduling algorithm allocates $c_2$ to $n_4$, after which the entries in the second and fourth columns (representing components to migrate \emph{to} nodes $n_2$ and $n_4$), and the fourth row (representing components to migrate \emph{out of} $n_2$ and $n_4$) are updated. All the entries in the second row are not updated because $c_2$ is not considered in the following scheduling. Let the migration threshold be $\varepsilon$=5ms, we can see that after this loop, the scheduling process is completed because no further effective migration can be conducted: all the values of entries in the updated matrix are smaller than 5ms.

The complexity of each scheduling interval is $O(m^2 \cdot k)$. Specifically, the performance matrix can be constructed in $O(m \cdot k)$ time. The scheduling process can be completed within $m$ loops, where each loops takes $O(m \cdot k)$ to find the optimal migration and $O(m+ m \cdot k)$ to update the matrix.

\section{Evaluation}  \label{Section: Evaluation}
\subsection{Experiment methodology} \label{Section: Experiment methodology}
\textbf{Experiment platform}. The experiments were conducted in a set of 30 nodes connected with a 1Gb ethernet network. Each node has two 6-core Xeon E5645 processors and hosts multiple VMs using Xen Virtual Machine Monitor (VMM). The operating system of both physical machines and VMs is SUSE Linux Enterprise Server (SLES)-11-SP1. The Xen, JDK versions are 4.0, 1.7.0, respectively. In addition, the versions of Nutch (search engine), Hadoop, and Spark are 1.1, 1.0.2, and 0.8.0, and the versions of Storm, Python, and Zookeeper are 0.9.2, 2.6, and 3.4.6, respectively.

%\begin{table}
%\center
%\begin{lrbox}{\tablebox}
%\begin{tabular}{|c|c|c|c|}
%  \hline
%  \multicolumn{2}{|c|}{\bf CPU type} & \multicolumn{2}{c|}{Intel \textregistered Xeon E5645} \\ \hline
%  \multicolumn{2}{|c|}{\bf Number of processors}  &\multicolumn{2}{c|}{2} \\ \hline
%  \multicolumn{2}{|c|}{\bf Number of cores / CPU}  &\multicolumn{2}{c|}{6 cores@2.40G} \\ \hline
% L1 DCache & L1 ICache & L2 Cache & L3 Cache \\ \hline
%6 $\times$ 32KB& 6 $\times$ 32KB&6 $\times$ 256KB& 12MB \\ \hline
%  \hline
%\end{tabular}
%\end{lrbox}
%\scalebox{0.8}{\usebox{\tablebox}}
%\caption{Node configuration details}\label{hwconfigeration}
%\end{table}

\textbf{Workloads}. We use representative workloads from the open-source BigDataBench workload suite \cite{opensourceBigDataBench}. The Nutch web search engine \cite{nutchsearch} represents the latency-critical online service and its online web search performance was tested. As shown in Figure \ref{Fig: SearchEngine}, this service has three stages and we call the components at Stage 1, 2, and 3 \emph{segmenting components}, \emph{searching components}, and \emph{aggregating components}, respectively.
The batch jobs involve a variety of Hadoop MapReduce and Spark jobs.
Hadoop jobs include the two typical CPU-intensive workloads with float point and integer calculations (Na\"{i}ve Bayes classification and WordCount) and one workload having similar demands for CPU and I/O resources (Page Index).
Spark jobs are mostly I/O-intensive workloads including Na\"{i}ve Bayes, WordCount and Sort.
These short-running batch jobs whose execution time ranges from a few seconds to several minutes represent a large fraction of jobs in today's data center workloads \cite{reiss2012heterogeneity,chen2012interactive}.

%To evaluate our approach under dynamic workloads, the input sizes of these jobs range from 50MB to 7GB.

%As shown in Figure \ref{Fig: SearchEngine}, the search engine has three stages, each with one or more component.

%We call the components at Stage 1, 2, and 3 \emph{segmenting components}, \emph{searching components}, and \emph{aggregating components}, respectively.

%(i.e. latency of searching indexes documents)
%We applied the Nutch crawler to collect pages from the Wikipedia site\footnote{http://www.wikipedia.org/} and constructed indexes, after which online web search performance (i.e. latency of searching indexes documents) was tested in experiment.

%Naive Bayes and CF workloads are available from Mahout\footnote{http://mahout.apache.org/} and other workloads are available from the BigDataBench\footnote{http://prof.ict.ac.cn/BigDataBench/}

%The component-node allocation in the scheduling of the service builds upon Storm, a popular distributed realtime system for data processing.

\textbf{Compared techniques}.
Two classes of state-of-the-art latency reduction techniques are compared. (i) \emph{Request redundancy} \cite{vulimiri2012more, ananthanarayanan2013effective,stewart2013zoolander}. For each request, multiple replicas are created for parallel execution and only the quickest replica is used. Two different redundancy policies, which generate three or five replicas were tested. (ii) \emph{Request reissue} \cite{jalaparti2013speeding,tailatScale}. A request is first sent to the most approximate component for execution, and a replica of this request is sent if the first one is not completed after a brief delay. The quickest replica is then used. Two reissue policies, which send a secondary request after the first has been executed for more than the 90th percentile or the 99th percentile of the expected latency for this class of requests, were tested.

For simplicity, we will call the four compared techniques, \emph{RED-3}, \emph{RED-5}, \emph{RI-90}, \emph{RI-99}. We also call the basic technique without any redundancy or reissue \emph{Basic}, and our predictive component-level scheduling approach \emph{PCS}.

\textbf{Metrics}. Two metrics are used to evaluate the performance of the search engine service. The \emph{first} metric is the 99th percentile latency of individual components of all requests. In the case of the request redundancy and reissue techniques, this metric denotes latencies of components belonging to the quickest replica. The \emph{second} metric is the average overall service latency of all requests.

%Given that 100ms is the typical required latency to deliver satisfactory service to end users \cite{tailatScale}, the \emph{third} metric is the percentage of requests whose overall latencies are larger than 100ms.

\textbf{Measurement method}. In the experiments, the monitor dynamically inspects the running service, including its request arrival rate and resource contention information listed in Table \ref{table: Usage information of shared resources}. The monitor obtains the request arrival rate and the system-level contention information once every second and the micro-architectural contention information once every minute. This measurement method guarantees low overheads in monitoring and does not affect the application performance.

% (e.g. the CPU usage is less than 1\%)
%At each scheduling interval, the proposed algorithm collects all the monitoring data and take the mean.

%A set of non-invasive and cost-effective online monitors were developed to dynamically extract the running service, including its request arrival rate and contention information of shared resources listed in Table \ref{table: Usage information of shared resources}.

\subsection{Prediction accuracy} \label{Section: Prediction accuracy}
The effectiveness of the proposed scheduling approach is considerably impacted by the performance model's accuracy. To evaluate this accuracy, we ran each searching component of the service on a VM with 1 core and 1GB memory, and used another VM with 4 core and 4GB memory co-located on the same node to run a Hadoop or Spark job of different input sizes. In each test, we trained the regression models based on the historical running information and predicted the component's service using the constructed models.

%collected the monitoring information to predict the component's latency using the performance model.
%Using the profiling information, regression models were trained.

%Using the microbenchmark tool introduced in Section \ref{Section: Basic performance model}, we generated contentions on each shared resource listed in Table \ref{table: Usage information of shared resources} and stepwise tuned up the contention intensity until the overall service latency is increased above an acceptable level (typically 5\% larger than 100ms).

%Ten contention intensities were tested for each resource type and the whole profiling was completed in 20 minutes.

%according to the AdaBoost algorithm and they were combined to produce the final regression model.

%By combining a small amount of profiling information from the workload itself with the large amount of data from previously-scheduled workloads

%To emulate real scenarios of co-running applications, we created two LinuX Containers (LXC) \cite{banga1999resource} to sandbox measurement runs. Each container limits its resource usage to 4 CPU cores and 4 GB memory via \emph{cgroups}. The other 16 CPU cores and 16 GB memory are shielded.

%several intensities?
%conducted resource contentions of different levels under the condition that the overall latency of the service varies is less than 100ms.

\begin{figure*}
\centering
  \includegraphics[scale=0.51]{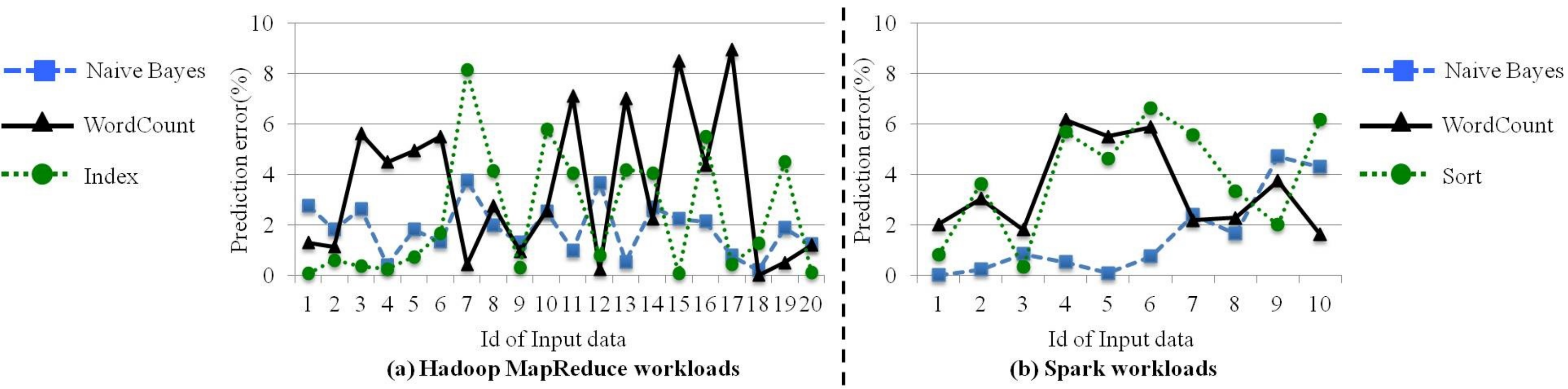}\\
  \caption{Prediction errors of the performance model under different performance interferences}
  \label{Fig: Prediction errors of the performance model under different performance interferences}
\end{figure*}

As listed in Figure \ref{Fig: Prediction errors of the performance model under different performance interferences}, in our evaluation, the Hadoop workloads have 20 different input sizes ranging from 50MB to 4GB, and the Spark workloads have 10 different input sizes ranging from 200MB to 7GB, thus having distinct performance interferences to the component's latency. As shown in Figure \ref{Fig: Prediction errors of the performance model under different performance interferences}, the prediction errors are \emph{smaller} than 3\%, 5\%, and 8\% in 63.33\%, 82.22\%, and 96.67\% of the evaluation cases, respectively. When considering all the input sizes, the average prediction error is 2.68\%, indicating the performance model keeps a good track of the observed latency and it is sufficient for our scheduling heuristic to achieve a near-optimal performance.

%Each input size was tested five times under the request arrival rates of 10 requests/second and we took the average.
%The prediction errors are mainly due to the measurement errors of resource contention information. In particular, the sampling rate of hardware performance counters (Perf and Oprofile) determines both the accuracy of micro-architectural information and the profiling costs: the higher the rate, the more accurate the information but the larger the cost. In this experiment, we obtained the micro-architectural contention information once each second, this guarantees the low overheads of monitoring (both the CPU and memory utilizations of our monitors are less than 1\%). Nevertheless, we believe that these are the reasonable trade-offs between the accuracy and the monitoring overheads, since the accuracy of the model is sufficient for our scheduling heuristic to achieve a near-optimal performance.

%This guarantee the information that drives scheduling decisions is accurate.

%\begin{table}[h!]
%  \caption{Average Prediction errors of the performance model}
%  \centering
%  \begin{tabular}{|l|l|l|}
%    \hline
%    \textbf{Workload} & \textbf{Input size ranges} & \textbf{Prediction error} \\
%    \hline
%    Na\"{i}ve Bayes (Hadoop) & 50MB to 5GB  & 1.98\%\\
%    \hline
%    WordCount (Hadoop) & 50MB to 2GB  & 5.11\%\\
%    \hline
%    Index (Hadoop) & 150MB to 4GB  & 2.23\%\\
%    \hline
%    Na\"{i}ve Bayes (Spark) & 100MB to 7GB  & 1.44\%\\
%    \hline
%    WordCount (Spark) & 50MB to 2GB  & 3.41\%\\
%    \hline
%    Sort (Spark) & 50MB to 2GB  & 3.88\%\\
%    \hline
%  \end{tabular}
%  \label{table: Prediction errors of the performance model}
%\end{table}

%> reasons:
%1> Storm itself,
%2> Although depends on, fixing the input size, the co-running, slightly changes the usage of dynamic set --> map/reduce

%--> refer to two papers
% We ensure that the maximum number of active client sessions is within the servers' maximum capabilities.

%\subsection{Mitigation of tail latency and overall latency}

%\textbf{Discussion the impact of machine heterogeneity}

\subsection{Service Performance} \label{Section: Service Performance}
\textbf{Evaluation setting}. Following the deployment settings of the previous section, we tested the performance of the Nutch search engine service whose searching components are deployed in 100 VMs. Each component co-locates with a mixed of batch jobs running on VMs of the same node. The Hadoop workloads were tested with continuously changing input data sizes ranging from 1MB to 10GB. Six request arrival rates, namely 10, 20, 50, 100, 200, 500 requests/second, were tested to compare the latency reduction techniques under online services' diurnal variation in load.

%The test of each arrival rate lasts 60 minutes and the scheduling interval is 10 minutes.

\textbf{Migration threshold}. As explained in Section \ref{Section: Algorithm}, the proposed scheduling algorithm employs a migration threshold to control latency reduction and throttle non-beneficial component migration. This threshold should be reasonably high to filter out most of the detrimental migrations whose overheads are larger than the possible latency reduction. On the other hand, the threshold cannot be too high to miss the opportunities for latency reduction.
The major overhead of migrating a component is caused by the movement from its current VM to the destination VM. The component runs on a VM installed Storm and its migration is enforced by calling Storm's deployment APIs. Specifically, Storm first uploads the source codes (e.g. codes for looking up indexes for documents) and the configuration information of the component to ZooKeeper \cite{ZooKeeperWebsite}, a widely used distributed coordination system to manage application deployment. ZooKeeper then allocates them to a new component on the destination VM. At each scheduling interval, the migration of components (e.g. 10 to 20 components) can be completed within 3 seconds without interrupting the running services and only causes small consumptions of memory and I/O resources.
Considering the migration cost, we find out that 5\% of the accepted overall service latency (100ms) is a reasonable threshold value for the studied online services and thus the threshold in scheduling is set as 5ms. Applying an adaptive threshold to improve the service performance is possible, but it is beyond the scope of this paper.

%In the \emph{second} strategy that directly conducts online migration of VMs between two nodes, our results show migrating one Xen VM hosting a component needs 60.89 seconds and the migration time gradually increases to 164.48 seconds when migrating eight such VMs.
%We tested two migration strategies. In the \emph{first} strategy,
%migrating eight such VMs needs 164.48 seconds.

%Given that 100ms is the typical required latency to deliver satisfactory service to end users \cite{tailatScale},

%Scheduling overheads: two aspects, resources (only new deployment plan and source codes of the component to process requests) + recovery time (percentages)

%\textbf{Impact of tail latency on service performance}.

\begin{figure*}
\centering
  \includegraphics[scale=0.58]{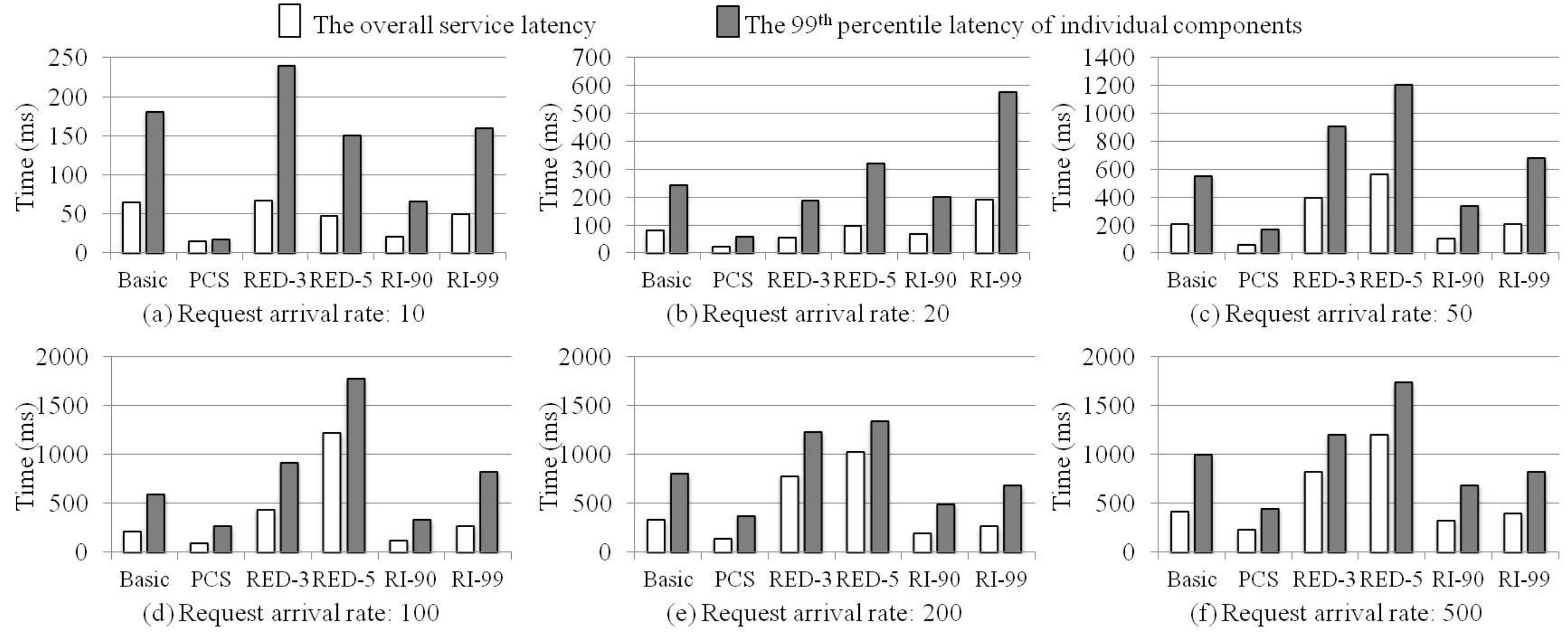}\\
  \caption{Comparison of overall service latency and the tail latency under different request arrival rates}
  \label{Fig: Overall service time and tail latency under different request arrival rates}
\end{figure*}

%\textbf{Comparison of different techniques}.
\textbf{Evaluation results}.
Figure \ref{Fig: Overall service time and tail latency under different request arrival rates} shows the comparison of service performance for six different techniques. The results show that PCS achieves the smallest tail latencies and the overall service latencies in all cases. This is because during the execution of the service, PCS dynamically enforces different component-node allocations along with the latest performance interference changes on different nodes and reduces the component latency variability by migrating the straggling components to nodes with less resource contentions.

By contrast, the \emph{request redundancy} technique just collects responses from the quickest component based on the current service deployment, missing the opportunity to migrate the components to the idlest nodes with the least performance interference. Figure \ref{Fig: Overall service time and tail latency under different request arrival rates}(a) and (b) show that this technique achieves some latency reduction under light workloads. However, when the arrival rate gradually increases to 500, Figure \ref{Fig: Overall service time and tail latency under different request arrival rates}(c) to (f) show that this technique adversely causes longer latencies compared to those of Basic. In particular, RED-5 causes the longest latencies because it produces the largest workloads, namely incurring the longest queueing delay, among all techniques.
Although the redundancy technique employs the cancelation mechanism that sends messages to cancel other queuing replicas when one replica begins execution, the components still execute replicas of the same request unnecessarily. This phenomenon mainly comes from two sources: (i) all replicas of a request are sent to multiple components simultaneously, hence two components having similar performances may start executing the requests in similar time; (ii) there is a network message delay for different components to communicate each other's status, hence two components may start executing the same request and the cancelation messages are both in the flight to each other.
Moreover, the \emph{request reissue} technique applies a conservative redundancy mechanism that only creates replicas for requests judged as outliers (i.e. requests whose execution time is larger than an expected latency). Results in Figure \ref{Fig: Overall service time and tail latency under different request arrival rates} show that compared to the request redundancy technique, this conservative reissue technique causes less performance deterioration when load becomes heavier.

\textbf{Results}. \emph{Considering all the six request arrival rates, PCS achieves 67.05\% reduction in the 99th component latency and 64.16\% reduction in the overall service latency when comparing to the request redundancy and reissue techniques.}

\subsection{Scalability of scheduling} \label{Section: Scalability of scheduling}

%\emph{Scalability of scheduling}.

In proposed scheduling heuristic, the used performance mode is constructed based on profiling of each component. That is, only one out of all homogeneous components needs to be profiled and thereby avoiding the scalability issue associated with the service profiling. For example, in the tested search engine service, only three components (segmenting, searching and aggregating) need to be profiled. Meanwhile, the proposed scheduling algorithm estimates the service performance by analyzing the resource contention information obtained from each component, and hence the \emph{analysis time} scales only linearly with the number of components. Another important aspect of the scalability of scheduling is to \emph{search} the appropriate component-code allocation, and the time complexity of this is $O(m^2\cdot k)$ when allocating $m$ components to $k$ nodes.

\begin{figure}
\centering
  \includegraphics[scale=0.50]{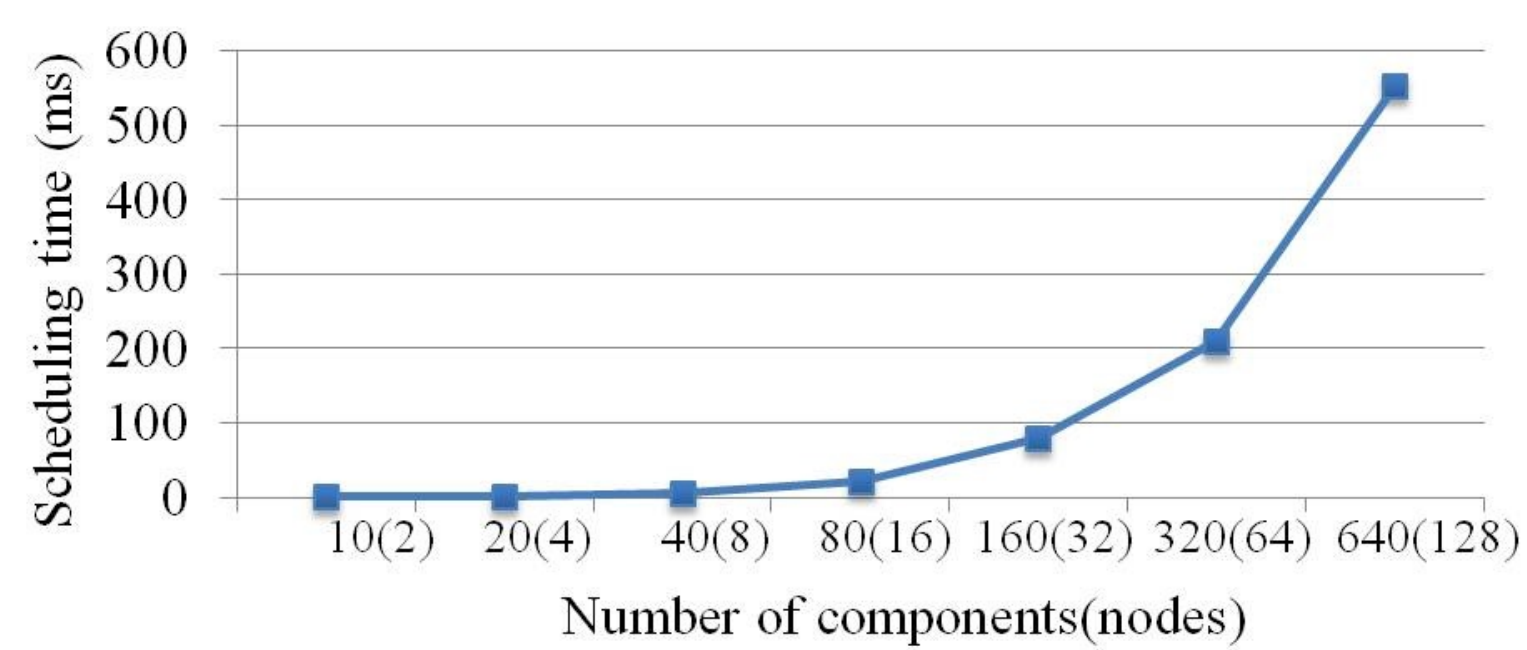}\\
  \caption{Scalability of the schduling algorithm}
  \label{Fig: Scalability of the searching algorithm}
\end{figure}

To evaluate the scheduling algorithm scalability, we measured both the analysis and searching time under different numbers of components and nodes. Figure \ref{Fig: Scalability of the searching algorithm} shows that even if the number of components reaches 640 (and the number of nodes reaches 128), the algorithm takes only 551ms to complete. This time is less than 0.1\% of the 600 seconds scheduling interval and hence can be ignored. For services with more components, the scheduler could apply a hierarchical strategy that divides the components into small groups of 640 components or less and finds the appropriate component-node allocation between groups and then within groups. The scheduling overhead therefore can remain low even with a large number of components.

\section{Related Work} \label{Section:Related Work}

\subsection{Application-level management of service performance}

At present, two categories of techniques have been proposed to meet the performance requirement of latency-critical services by alleviate the performance degradation due to resource sharing and contention. The \emph{first category} of techniques disallow the co-location of services with applications incurring large contention of resources such as caches \cite{kasture2014ubik} and CPU resources \cite{xu2013bobtail}.
The \emph{second category} of techniques dynamically manage applications to meet their performance requirement at run-time according to the monitored interference metrics, such as the LLC miss rate reflecting cache contentions \cite{ahn2012dynamic} and the bandwidths reflecting I/O resource contentions \cite{xu2010mitigating}.
These techniques focus on addressing performance variability of applications by viewing the application as a whole, ignoring issues relating to fine-grained latency variability of its individual components. However, these components' tail latency dominates performance of large-scale, parallel services.

\subsection{Tail latency reduction techniques}

We now review four categories of reduction techniques.

\textbf{Modifying hardware/software systems}. These techniques aim at solving the tail latency caused by system design issues. Those include architecture-level design that disables the power saving model to promote system performance \cite{wang2013impact}; OS-level design that changes the default kernel scheduler to a better scheduler (e.g. Borrowed Virtual Time (BVT)) with better support for time-sensitive requests \cite{leverich2014reconciling}.

\textbf{Adding additional resources}. These techniques require additional resources to handle slow requests, either by increasing the parallelism degree of the request processing \cite{jeon2014predictive} or adding new server components \cite{stewart2013zoolander}.

\textbf{Partially processing request}. These techniques reduce tail latency by only using a portion (e.g. 90\%) of the quickest sub-requests \cite{jalaparti2013speeding} or a synopsis representing the entire input data at a high level of approximation \cite{han2015sarp}, thus scarifying result correctness such as query accuracy for reducing service latency.

The approach proposed in this work can work together with the above techniques to reduce tail latency, thus forming a complement to these techniques.
Both this work and the fourth category of techniques, namely \textbf{request redundancy} \cite{vulimiri2012more, ananthanarayanan2013effective,stewart2013zoolander} \textbf{and reissue} \cite{jalaparti2013speeding,tailatScale} explained in Section \ref{Section: Experiment methodology}, reduce tail latency by addressing component latency variability. The key idea of the request redundancy technique is to execute the same request on multiple components so as to reduce its latency by using the quickest one. Although these techniques work well when workloads underutilize system resources \cite{stewart2013zoolander}, they start hurting the service performance and adversely worsen the latency when load gets heavier \cite{shah2013redundant} .

\section{Conclusion} \label{Section: Conclusion}
This paper presents a component-level scheduling framework that can dynamically schedule components of a service across hundreds of machines in a cloud data center. To adapt to the changing performance interferences and workloads, this framework leverages cost-efficient online monitors and an analytic performance model to simultaneously predict the components' latency when running on different nodes. Using the predicted performance, the scheduler identifies straggling components and enforces near optimum component-node allocations. By comparing to the best well-known techniques on reducing tail latency, we demonstrate that our approach achieves significant reductions in both component tail latency and overall service latency.



%\bibliographystyle{IEEEtran}
%\begin{thebibliography}{1}


%\end{thebibliography}

\section{Acknowledgements}
We sincerely thank Moustafa M. Ghanem and Li Guo and for their useful comments, and the anonymous reviewers for their feedback on earlier versions of this manuscript. This work is supported by Chinese 973 projects under Grants No. 2014CB340402.

\bibliographystyle{plain}
\bibliography{references}
\end{document}